\documentclass{article}
\usepackage{graphicx}
\title{Nonextensive Entropy, Prior PDFs and Spontaneous Symmetry Breaking}
\author{Fariel Shafee\\Department of Physics\\ Princeton University\\Princeton, NJ 08540}
\begin{document}
\date{}
\maketitle
\begin{abstract}
We show that using nonextensive entropy can lead to spontaneous symmetry breaking when a parameter changes its value from that applicable for a symmetric domain, as in field theory. We give the physical reasons and also show that even for symmetric Dirichlet priors, such a definition of the entropy and the parameter value can lead to asymmetry when entropy is maximized.
\end{abstract}

\section{Introduction}

Nonextensive entropies, such as that defined by Tsallis\cite{tsallis}, or more recently by us\cite{ours}, among others\cite{kan1,kan2} differ from the conventional Boltzmann-Shannon form, which is extensive in the sense of being additive when two subsystems in equilibrium are joined together. In nonextensive forms the combined value of entropy may be, in general, higher or lower than the sum of the entropies for the subunits joined. The deviation is , therefore, ascribable to interactions of a nonrandom nature among the microsystems comprising each subunit.

The maximum value of extensive entropy occurs when the probabilities are equally distributed among all the possible states of the system. In other words the conventional entropy is maximal for the most symmetric distribution of the microsystems. In the present paper we show that for nonextensive entropies defined on terms of phase cell deformations, the maximal entropy may not correspond to an equidistribution of probability among the states.

\section{Nonextensive Entropy}

The classical entropy

\begin{equation}\label{shannon}
S = -\sum_i p_i \log p_i
\end{equation}
may be modified in several ways. The well-known Tsallis form generalizes the logarithm:

\begin{equation}
\log p \rightarrow {\bf Log}_q p \equiv \frac{1-p^q}{1-q}
\end{equation}

For our entropy we make the measure a fractal:
\begin{equation}\label{ours}
S = - \sum_i p_i^q \log p_i
\end{equation}
with $q=1$ giving the classical Shannon entropy, as in the Tsallis case. In ref. \cite{ours} we have given detailed account of the different physical considerations that lead to our expression and also comparison of the statistical mechanical properties of the three entropies.

The justification of choosing Shannon or any other more generalized entropy, such as that of Tsallis or Renyi, or
the one we have presented elsewhere\cite{ours,entent}, lies eventually in the relevance or ``good fit'' such an entropy would produce in the data corresponding to a situation where the presence or lack of interactions among the members or other considerations suggest the need for a proper choice. However, data are always finite, and probability distribution is the limit of relative frequencies with an infinite sample. One, therefore faces the problem of
estimating the best PDF from a finite sample \cite{BCS1}. This PDF may be subject to the constraint of a known
entropy, in whatever way defined, as a functional of the PDF.

Mathematically, the problem of determining the best posterior PDF, given a rough prior PDF and data points, is
expressed formally by Bayes' Theorem. However, the constraint of the constant entropy makes the functional integral impossible to handle even for a fairly simple prior as found by Wolpert and Wolf \cite{WW1} and by Nemenman, Shafee and Bialek\cite{NSB1}. The integrals involved were first considered in a general context by \cite{WW1}, and the question of priors was addressed in \cite{BCS1,NSB1}. It was discovered that, though the integral for the posterior was intractable, the moments of the entropy could be calculated with relative ease.

In \cite{NSB1} it has also been shown that for Dirichlet type priors \cite{dirichlet}

\begin{equation}\label{dir1}
P(p_i)= \prod p_i^{\beta}
\end{equation}
in particular (which give nice analytic moments with exact integrals, and hence, are hard to ignore) the Shannon
entropy is fixed by the exponent $\beta$ of the probabilities chosen for small data samples, and hence, not much information is obtained for  unusual distributions,such as that of Zipf, i.e. a prior has to be wisely guessed for any meaningful outcome. As a discrete set of bins has no metric, or even useful topology that can be made use of in Occam razor type of smoothing, in this paper other tricks were suggested to overcome the insensitiveness of the entropy.

We have noted already that the PDF associated with our proposed entropy differs from that of the Shannon entropy
by only a power of $p_i$, but this changes the symmetry of the integrations for the moments for the different terms for different bins. We,therefore, shall examine in this chapter if the nature of the moments are sufficiently changed by our entropy to indicate cases where data can pick this entropy in preference to Shannon or other entropies.

\section{Priors and Moments of Entropy}

For completeness, we mention here the formalism developed by Wolpert and Wolf \cite{WW1}. The uniform PDF is given by

 \begin{eqnarray}
  {\mathcal P}_{\rm unif}(\{p_i\}) = {1\over Z_{\rm unif}}\,\delta\left(
    1 - \sum_{i=1}^K p_i\right) \nonumber\\
     Z_{\rm unif} = \int_{\mathcal A}dp_1 dp_2 \cdots  dq_K
  \,\delta\left( 1 - \sum_{i=1}^K p_i\right)
\end{eqnarray}
where the $\delta$ function is for normalization of probabilities, $Z_{\rm unif}$ is the total volume occupied by
all models. The integration domain ${\mathcal V}$ is bounded by each $p_i$  in the range $[0,1]$. Because of the normalization constraint, any specific set of $\{p_i\}$ chosen from this distribution is not uniformly distributed and ``uniformity'' means simply that all distributions that obey the normalization constraint are equally likely {\em a priori}.

We can find  the probability of the model $\{ p_i\}$ with Bayes rule as
\begin{eqnarray}
  P(\{ p_i\}| \{ n_i\} ) = \frac{P(\{ n_i\} | \{ p_i\})
    {\mathcal P}_{\rm unif}(\{p_i\})}{P_{\rm unif}(\{ n_i\})} \nonumber \\
  P(\{ n_i\} | \{ p_i\}) = \prod_{i=1}^K (p_i)^{n_i}.
\end{eqnarray}

Generalizing these ideas, we have considered priors with a power-law dependence on the probabilities calculated
as

\begin{equation}
{\mathcal P}_\beta(\{p_i\}) = {1\over Z(\beta)}
\delta\left( 1 - \sum_{i=1}^K p_i\right)
\prod_{i=1}^K p_i^{\beta-1} \,,
\label{P(p)}
\end{equation}
 It has been shown \cite{NSB1} that if $p_i$'s are
generated in sequence [ $i=1 \rightarrow K$] from the Beta--distribution
\begin{eqnarray}
  P(p_i) = B\left(\frac{q_i}{1-\sum_{j<i} p_j}; \beta, (K-i)\beta
  \right) \nonumber \\
  B\left(x; a,b \right) =\frac{x^{a-1}(1-x)^{b-1}}{B(a,b)}
\end{eqnarray}
gives  the probability of the whole sequence $\{p_i\}$ as ${\mathcal P}_{\beta}(\{p_i\})$.

Random simulation of PDF's with different shapes (a few bins occupied, versus more spread out ones) show that the
entropy depends largely on the parameter $\beta$ of the prior and hence, sparse data has virtually no role in getting the output distribution shape. This would seem unsatisfactory, and some adjustments appear to be needed to get any useful information out.

We shall not here repeat the methods and results of \cite{NSB1}, which considers only Shannon entropy.

\section{Comparison of Shannon and Our Entropy}

In our case with the entropy function given by Eqn. ~\ref{ours}, we note that it does not involve a simple replacement of the exponents $n_i$ of $p_i$ by $n_i+q-1$ in the case of the Dirichlet prior (Eqn.~\ref{dir1}) in the product involved in the moment determination
integrals given in \cite{WW1}, but a complete re-calculation of the moment, using the same techniques given in \cite{WW1}. Apparently,  the maximal value of entropy should correspond to the most flat distribution, i.e.

\begin{equation} \label{ourmaxent}
S_{max} =  K^{(1-q)}\log(K)
\end{equation}

In the limit of extremely sparse, nearly zero data ($n_i=0$), we get for the first moment, i.e. the expected entropy,

\begin{equation}\label{ourmom1}
\langle S_1 \rangle/ \langle S_0 \rangle = K \frac { \Gamma (\beta +q )}{\Gamma (\beta)}
\frac{\Gamma (\beta K)}{\Gamma( \beta K + q)} \Delta \Phi^0 (\beta K+q, \beta+q)
\end{equation}
where we have for conciseness used the notation of ref. \cite{WW1}

\begin{equation}\label{psinotation}
\Delta\Phi^p(a,b) = \Psi^{(p-1)}(a) - \Psi^{(p-1)}(b)
\end{equation}
$\Psi^n(x)$ being the polygamma function of order $n$ of the argument $x$. It can be checked easily that this
expression reduces to that in ref. \cite{NSB1} when $q=1$, i.e. when we use Shannon entropy.

\section{Results for Mean Entropy}

\begin{figure}[ht!]
\begin{center}
\includegraphics [width=8cm]{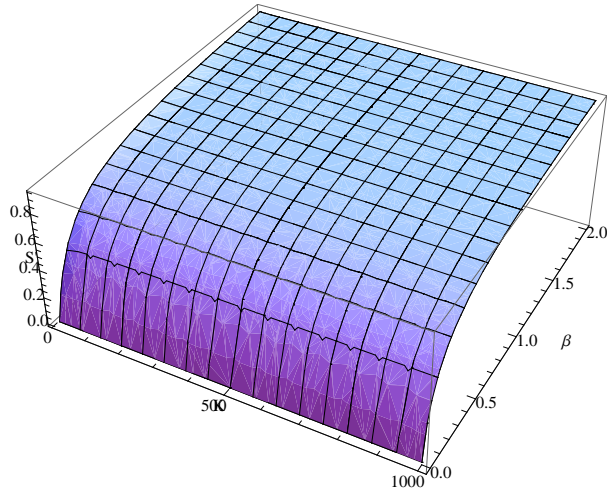}
\end{center}
 \caption{\label{myprior1} Ratio of expected value (first moment) of new entropy plotted against bin number $K$
 and prior exponent $\beta$
 for entropy parameter  $q=0.5$.}
\end{figure}

\begin{figure}[ht!]
\begin{center}
\includegraphics[width=8cm]{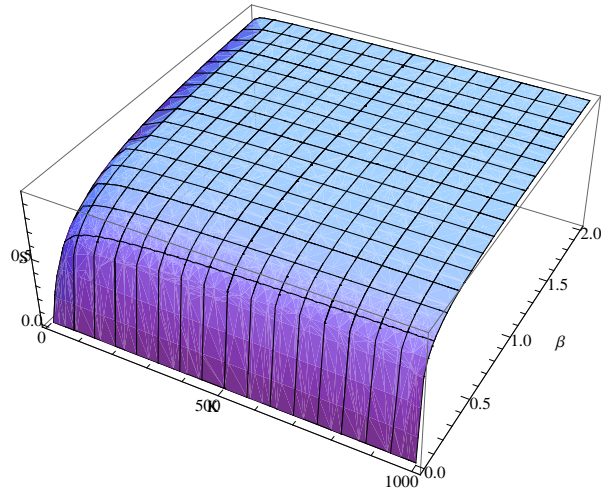}
\end{center}
 \caption{\label{myprior2}Same as Fig. ~\ref{myprior1}, but for $q=1.0$, i.e. Shannon entropy.}
\end{figure}

\begin{figure}[ht!]
\begin{center}
\includegraphics [width=8cm]{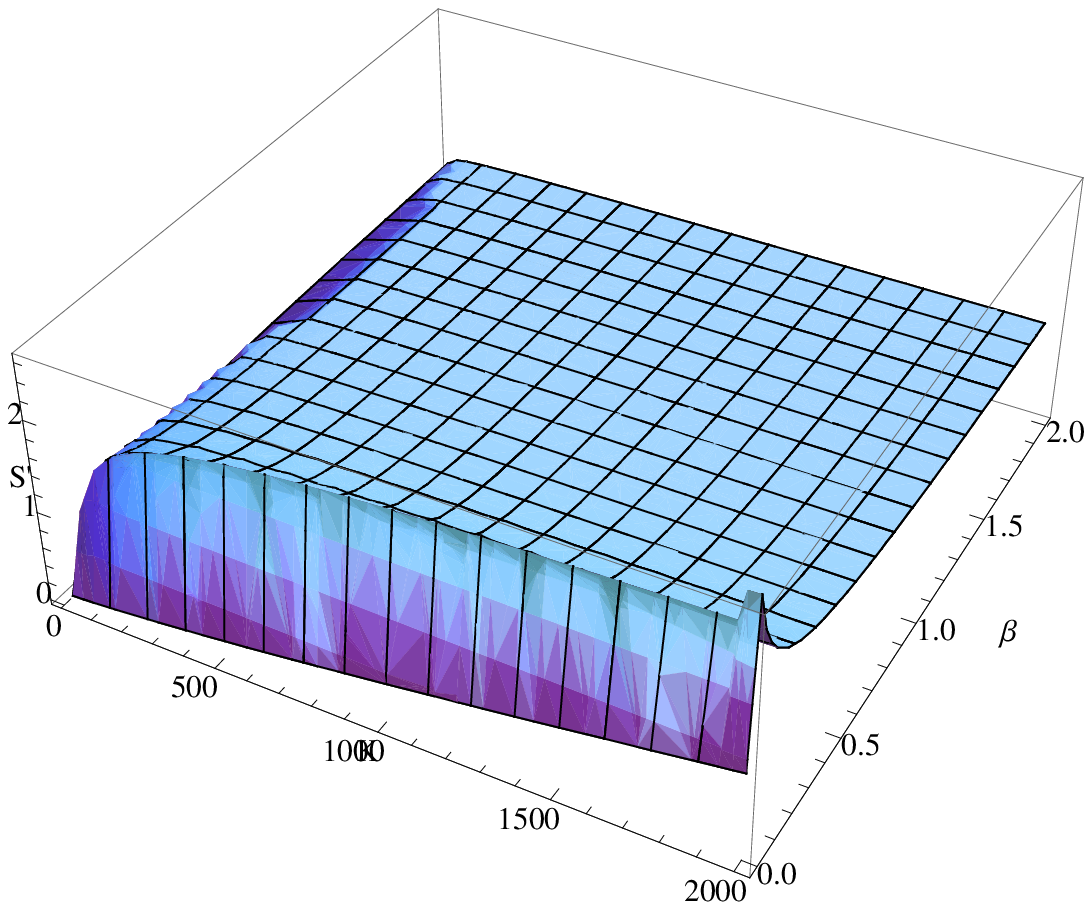}
\end{center}
 \caption{\label{myprior3}As previous two figures, but for $q=1.5$.}
\end{figure}

So, we now have, unlike Shannon, a parameter $q$ that may produce the difference from the Shannon case, where $q$
is fixed at unity. In Figs.~\ref{myprior1} -~\ref{myprior3} we show the variation of the ratio of $\langle S_1 \rangle/S_{max}$ with variable bin number $K$.
In ref. \cite{NSB1} we have commented how insensitive the Dirichlet prior \cite{dirichlet} is when Shannon entropy is
considered in the straightforward manner given in ref. \cite{WW1}. In our generalized form of the entropy, we note that by changing the parameter, specific to our form of the entropy, for $q > 1$, we get a peak for small $\beta$ and large $K$ values.

This peak allows us to choose uniform Dirichlet priors with appropriate $q$ value, that would nevertheless lead to
asymmetry not possible with Shannon entropy. In other words, instead of the priors, we can feed the information about expected asymmetry of the PDF to the entropy
with no need to choose particular bins. The nonextensivity of our entropy, coming possibly from interaction among
the units, gives rise to situations where the entropy maxima do not increase with the number of bins like $\log(K)$,  but being $K^{(1-q)}\log(K)$, may be extended or squeezed, according to the value of $q$ being less than or greater than unity.

\section{Spontaneous Symmetry Breaking}

 The interesting thing to note is that for $q>1$ and large $K$, at small prioric parameter $\beta$, the entropy
 peak exceeds  the normally expected expression in Eqn. ~\ref{ourmaxent}, with full $K$,  so, the expected value of entropy is  seen to exceed  the formal maximum. The clustering or repulsive effects,  change the measure of disorder from the Shannon type  entropy. So, the  highest expected value of entropy may correspond not to a uniformly distributed population, but to that  corresponding to one  with a smaller subset that is populated.  This means that for our entropy the most uniform distribution is not  the least informative, the $p^q$ weighting distorts it to an uneven distribution for the expected maximal entropy value.
 This result is in some ways similar to spontaneous symmetry breaking in field theory, where the variation of a
 parameter leads to broken-symmetry  energy minima.

\begin{figure}[ht!]
\begin{center}
\includegraphics [width=8cm]{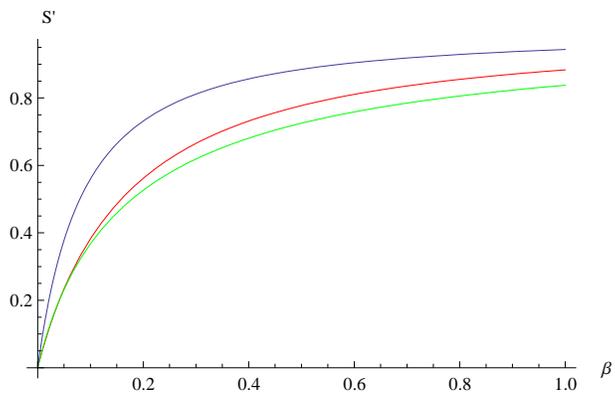}
\end{center}
 \caption{\label{myprior4} Clearer view in 2-dimensional plot, with $K=10$. Red, green and blue lines are for
 $q=0.5,1.0$ and $1.5$
 respectively}
\end{figure}

\begin{figure}[ht!]
\begin{center}
\includegraphics [width=8cm]{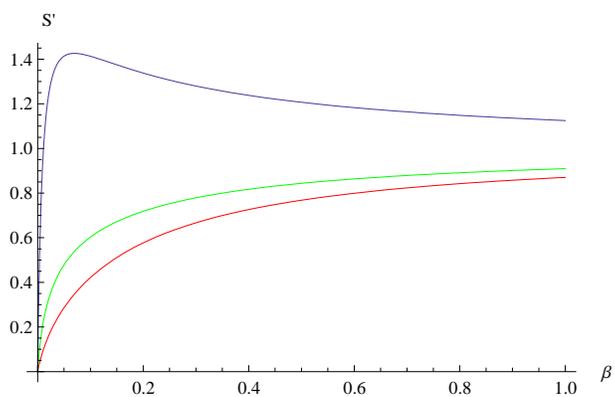}
\end{center}
 \caption{\label{myprior5}As Fig. ~\ref{myprior4}, but for bin number $K=100$. }
\end{figure}

\begin{figure}[ht!]
\begin{center}
\includegraphics [width=8cm]{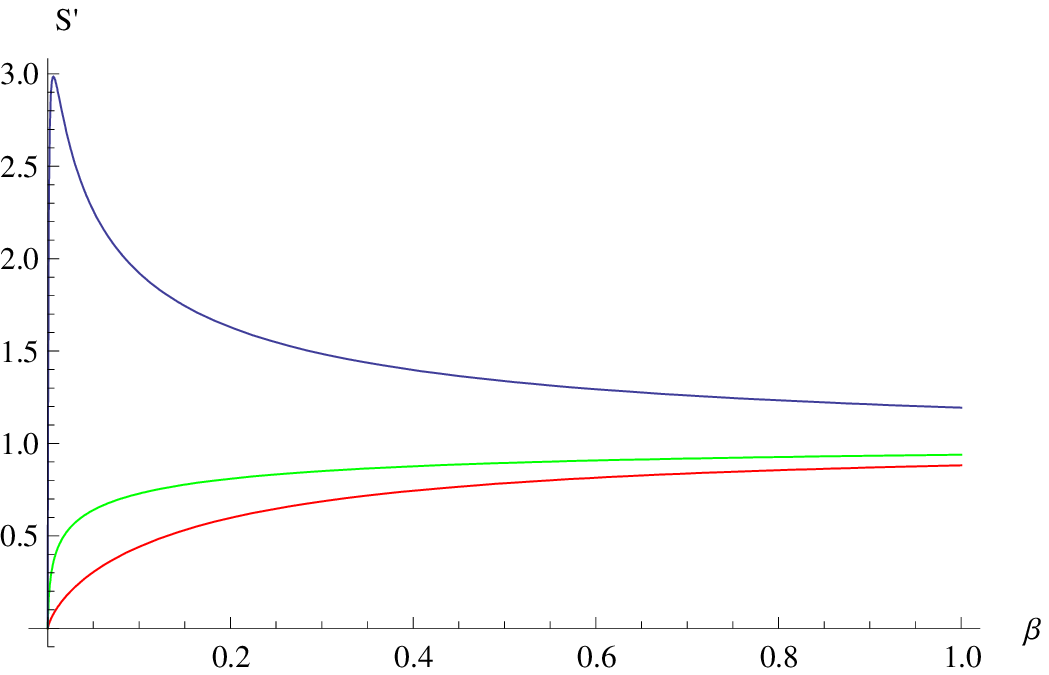}
\end{center}
 \caption{\label{myprior6}As previous two figures, but for $K=1000$.}
\end{figure}

A neater view of these results can be seen in Figs. ~\ref{myprior4}-~\ref{myprior6} with $K$ values fixed.

We have not obtained the second moment, i.e. the standard deviation , or the spread, of the entropy distribution,
because, with our entropy and an arbitrary $q$, the expressions cannot be obtained in the simple form of ref. ~\cite{NSB1}. We, can however, expect that the variation of the higher moments from the Shannon case will be less than the first moment, because higher derivatives of the $\Gamma$ functions are
smoother. We shall assume the spreads are narrow enough to concentrate on the first moments only.

Apart from the PDF estimates above, this picture of broken symmetry for the maximal entropy when the parameter
$q>1$ is also manifest directly in an explicit calculation of the entropy using our prescription with a simple
three-state system. The symmetric expected maximal entropy in this case should be
\begin{equation}\label{ssb1}
S_{max}= - 3 p^q \log p
\end{equation}
with $p=1/3$.

\begin{figure}[ht!]
\begin{center}
\includegraphics [width=8cm]{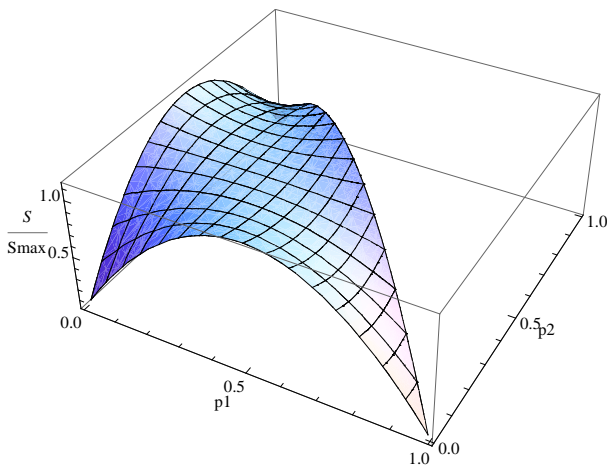}
\end{center}
 \caption{\label{figssb1} Our entropy for a three-state system, with parameter $q=2.44$, as two independent
 probabilities $p_1$ and $p_2$ are varied wit the constraint $p_1+p_2+p_3=1$. The expected maximum at the symmetry
 point $p_1=p_2=p_3$ turns out to be a local maximum. The global maxima are not at the end points with one of the
 probabilities going up to unity and the others vanishing, which gives zero entropy as expected, but occurs near
 such end points, as shown clearly in the next figure.}
\end{figure}

\begin{figure}[ht!]

\begin{center}
\includegraphics [width=8cm]{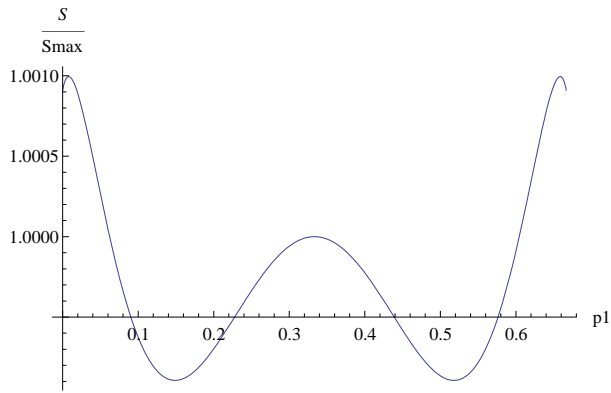}
\end{center}
 \caption{\label{figssb2}Two-dimensional version of the previous Fig. ~\ref{figssb1}, with $p_1=1/3$ fixed, so that only
 $p_2$ varies. This shows a clearer picture of local maximum at the symmetry point and global maxima near the end
 points.}
\end{figure}

With two of the probabilities $p_1$ and $p_2$ running free from $0$ to $1$ with the constraint $p_1+p_2+p_3=1$,
the plot for the entropy

\begin{equation}\label{ssb2}
S= - p_1^q \log p_1 - p_2^q \log p_2 - (1-p_1-p_2) \log(1-p_1-p_2)
\end{equation}
we plot $S/S_{max}$ in Figs.~\ref{figssb1},~\ref{figssb2}. For $q=2.44$ we obtain the most interesting behavior, with a local maximum at the point of symmetry $p_1=p_2=p_3=1/3$, which is not the global maximum. For $q\leq 1$ the symmetry
point gives the global maximum.

A physical explanation of the SSB may be the distortion introduced by nonrandom interactions in the volumes of the 'phase space' of the states. In ref. \cite{ours} we have shown how the new entropy is related to such volumes, in terms of Shannon coding theorem. Apparently this distortion introduces a mixing of states that reduces the weights of clearly defined states and hence introduces a new measure of uncertainty not present in the case of Shannon entropy. As a result it is entropically preferable to leave some states underpopulated to increase the total entropy by overpopulating others. In other words we have a reduction of the problem from $N$ states to less, but with a measure factor with less dimunition that overcompensates the decrease in the logarithmic factor. In a field-theoretic model with the Lagrangain
\begin{eqnarray}
L= |\partial \phi|^2 -V_I  \nonumber \\
V_I = -\mu^2 |\phi|^2 + \lambda |\phi|^4
\end{eqnarray}
for $\mu^2$ and $\lambda$ with opposite signs, the lowest energy state is the symmetric vacuum (no state occupied), and for same sign the vacuum becomes a local maximum, with a ring of  minima at $|\phi|= |\mu/\surd{(2\lambda)}|$, which forces us to choose a unique vacuum with a particular complex $\phi$ having this magnitude. In the case of entropy, for $q\leq1$ we have the highest entropy for all states equally populated, and for $q>1$ the configuration with a symmetric flat PDF is no longer the one with the highest entropy.

\section{Conclusions}

We have seen that average entropies corresponding to uniformly symmetric Dirichlet type priors can be obtained exactly even for nonextensive entropies of a type we have described earlier. Remarkably this entropy shows maxima for asymmetric probability distributions, which can be considerably higher than the symmetric distribution,  unlike Shannon entropy. We think this asymmetry is a consequence of the distortion of the `phase cells' associated with the states, which may in turn be due to nonrandom interactions.

The author thanks Prof. Phil Broadbridge of the Australian Mathematical Sciences Institute for encouragement.


\begin{thebibliography}{99}

\bibitem{tsallis} C. Tsallis, Possible generalization of Boltzmann-Gibbs statistics, {\em J. Stat.Phys.}  {\bf52},
    479--487 (1988).

\bibitem{ours}F. Shafee, Lambert function and a new non-extensive form of entropy, {\em IMA Journal of Applied
    Mathematics}  {\bf72}, 785--800 (2007)

\bibitem{kan1} G. Kaniadakis,  Nonlinear kinetics underlying generalized
statistics. {\it Physica A} {\bf 296}, 405--425 (2001).

\bibitem{kan2} G. Kaniadakis,  Statistical mechanics in the context of special relativity. {\it Phys. Rev. E} {\bf
    66}, 056125 (2002).



\bibitem{NSB1}I.~Nemenman, F.~Shafee and W.~Bialek, Entropy and Inference, Revisited, in {\em Adv. Neur. Info.
    Processing  14}, eds. T.G.
    Dietterich, S. Becker and Z. Ghahramani (MIT Press, Cambridge, 2002) pp. 471--478.


\bibitem{entent} F. Shafee, Generalized Entropy with Clustering and Quantum Entangled States.
 cond-mat/0410554  (accepted by {\em Chaos, Solitons and Fractals})(2008)

\bibitem{BCS1} W.~Bialek, C.G.~ Callan and S.P.~Strong, Field theories for learning probability
distributions.{\em Phys. Rev. Lett.}  {\bf 77}, 4693--4697 (1996)

\bibitem{WW1}D.~Wolpert and D.~Wolf, Estimating functions of  probability distributions from a finite set of
    samples,  {\it Phys.~Rev.~E},
    {\bf 52}, 6841--6854 (1995).

\bibitem{dirichlet}E.T. Jaynes, "Monkeys, Kangaroos, and N", University of Cambridge Physics Dept. Report 1189
    (1984).
\end{thebibliography}
\end{document}